# Dirac's hole theory and the Pauli principle: clearing up the confusion.

Dan Solomon
Rauland-Borg Corporation
1802 W. Central Road
Mount Prospect, IL 60056
USA

Email: dan.solomon@rauland.com

**Abstract.**
In Dirac's hole theory (HT) the vacuum state is generally believed to be the state of minimum energy due to the assumption that the Pauli Exclusion Principle prevents the decay of positive energy electrons into occupied negative energy states. However recently papers have appeared that claim to show that there exist states with less energy than that of the vacuum[1][2][3]. Here we will consider a simple model of HT consisting of zero mass electrons in 1-1D space-time. It will be shown that for this model there are states with less energy than the HT vacuum state and that the Pauli Principle is obeyed. Therefore the conjecture that the Pauli Principle prevents the existence of states with less energy than the vacuum state is not correct.

## 1. Introduction.

It is well known that there are both positive and negative energy solutions to the Dirac equation. This creates a problem in that an electron in a positive energy state will quickly decay into a negative energy state in the presence of perturbations. This, of course, is not normally observed to occur. This problem is presumably resolved in Dirac's hole theory (HT) by assuming that all the negative energy states are occupied by a single electron and then evoking the Pauli exclusion principle to prevent the decay of positive energy electrons into negative energy states.

The proposition that the negative energy states are all occupied turns a one electron theory into an N-electron theory where $N \to \infty$. Due to the fact that the negative energy vacuum electrons obey the same Dirac equation as the positive energy electrons we have to, in principle, track the time evolution of an infinite number of states. Also the vacuum electrons in their unperturbed state are unobservable. All we can do is observe the differences from the unperturbed vacuum state.

It is generally assumed that the HT vacuum state is the state of minimum energy. That is, the energy of all other states must be greater than that of the vacuum state. However a number of papers by the author have shown this is not the case ([1][2][3]). It was shown in these papers that states exist in HT that have less energy than the vacuum state.

One possible objection to this result is that it seems to contradict the Pauli exclusion principle. It is the purpose of this paper to show that this is not the case. It will be shown that the existence of states with less energy than the HT vacuum state is perfectly consistent with the Pauli principle.

In this paper we will consider a "simple" quantum theory consisting of non-interacting zero mass electrons in a background classical electric field. The advantage of such a simple system is that we can easily obtain exact solutions for the Dirac equation for any arbitrary electric potential. This considerably simplifies the analysis.

In the following discussion we will assume that the system is in some initial state which consists of an infinite number of electrons occupying the negative energy states (the Dirac sea) along with a single positive energy electron. The system will be then perturbed by an electric field. Each electron will evolve in time according to the Dirac equation. After some period of time the electric field is removed and the change in the energy of each electron can be calculated. The total change in the energy of the system is the sum of these changes. It will be shown that it is possible to specify an electric field so that final energy is less than the energy of the vacuum state. It will be also shown that this result is entirely consistent with the Pauli exclusion principle.

## 2. The Dirac Equation

In order to simplify the discussion and avoid unnecessary mathematical details we will assume that the electrons have zero mass and are non-interacting, i.e., they only interact with an external electric potential. Also we will work in 1-1 dimensional space-time where the space dimension is taken along the z-axis and use natural units so that $\hbar = c = 1$. In this case the Dirac equation for a single electron in the presence of an external electric potential is,

$$i\frac{\partial \psi(z,t)}{\partial t} = H\psi(z,t) \tag{2.1}$$

where the Dirac Hamiltonian is given by,

$$H = H_0 + qV(z,t) \tag{2.2}$$

where $H_0$ is the Hamiltonian in the absence of interactions, $V(z,t)$ is an external electrical potential, and q is the electric charge. For zero mass electrons the free field Hamiltonian is given as,

$$H_0 = -i\sigma_3 \frac{\partial}{\partial z} \tag{2.3}$$

where $\sigma_3$ is the Pauli matrix with $\sigma_3 = \begin{pmatrix} 1 & 0 \\ 0 & -1 \end{pmatrix}$.

The solution of (2.1) can be easily shown to be,

$$\psi(z,t) = W(z,t)\psi_0(z,t) \tag{2.4}$$

where $\psi_0(z,t)$ is the solution to the free field equation,

$$i\frac{\partial \psi_0(z,t)}{\partial t} = H_0\psi_0(z,t) \tag{2.5}$$

and can be written as,

$$\psi_0(z,t) = e^{-iH_0 t}\psi_0(z) \tag{2.6}$$

The quantity $W(z,t)$ is given by,

$$W(z,t) = \begin{pmatrix} e^{-ic_1} & 0 \\ 0 & e^{-ic_2} \end{pmatrix} \tag{2.7}$$

where $c_1(z,t)$ and $c_2(z,t)$ satisfy the following differential equations,

$$\frac{\partial c_1}{\partial t} + \frac{\partial c_1}{\partial z} = qV \tag{2.8}$$

and,

$$\frac{\partial c_2}{\partial t} - \frac{\partial c_2}{\partial z} = qV \tag{2.9}$$

These relationships also satisfy,

$$\frac{\partial(c_1+c_2)}{\partial t} + \frac{\partial(c_1-c_2)}{\partial z} = 2qV \; ; \; \frac{\partial(c_1-c_2)}{\partial t} + \frac{\partial(c_1+c_2)}{\partial z} = 0 \tag{2.10}$$

Let $\varphi_{\lambda,p}^{(0)}(z)$ be the eigenfunctions of the free field Hamiltonian with energy eigenvalue $\varepsilon_{\lambda,p}^{(0)}$. They satisfy the relationship,

$$H_0 \varphi_{\lambda,p}^{(0)}(z) = \varepsilon_{\lambda,p}^{(0)} \varphi_{\lambda,p}^{(0)}(z) \tag{2.11}$$

where,

$$\varphi_{\lambda,p}^{(0)}(z) = \frac{1}{2\sqrt{L}} \begin{pmatrix} 1 + \frac{\lambda p}{|p|} \\ 1 - \frac{\lambda p}{|p|} \end{pmatrix} e^{ipz} \; ; \; \varepsilon_{\lambda,p}^{(0)} = \lambda|p| \tag{2.12}$$

and where $\lambda = \pm 1$ is the sign of the energy, $p$ is the momentum, and $L$ is the 1 dimensional integration volume. We assume periodic boundary conditions so that the momentum $p = 2\pi r/L$ where $r$ is an integer. According to the above definitions the quantities $\varphi_{-1,p}^{(0)}(z)$ are negative energy states with energy $\varepsilon_{-1,p}^{(0)} = -|p|$ and the quantities $\varphi_{+1,p}^{(0)}(z)$ are positive energy states with energy $\varepsilon_{+1,p}^{(0)} = |p|$.

The $\varphi_{\lambda,p}^{(0)}(z)$ form an orthonormal basis set and satisfy,

$$\int \varphi_{\lambda,p}^{(0)\dagger}(z)\varphi_{\lambda',p'}^{(0)}(z)dz = \delta_{\lambda\lambda'}\delta_{pp'} \tag{2.13}$$

where integration from $-L/2$ to $+L/2$ is implied. If the electric potential is zero then the $\varphi_{\lambda,p}^{(0)}(z)$ evolve in time according to,

$$\varphi_{\lambda,p}^{(0)}(z,t) = e^{-iH_0 t}\varphi_{\lambda,p}^{(0)}(z) = e^{-i\lambda|p|t}\varphi_{\lambda,p}^{(0)}(z) \tag{2.14}$$

The energy of a normalized wave function $\psi(z,t)$ is given by,

$$\xi(\psi(z,t)) = \int \psi^\dagger(z,t)\left(H_0 + V(z,t)\right)\psi(z,t)dz \tag{2.15}$$

In the case where $V$ is zero the energy equals the free field energy which is given by,

$$\xi_0(\psi(z,t)) = \int \psi^\dagger(z,t) H_0 \psi(z,t)dz \tag{2.16}$$

Now suppose at $t=0$ a normalized wave function $\psi_0(z)$ is specified and the electric potential is zero. Now apply an electric potential and then remove it at some future time $t_f$. The wave function has evolved into the state $\psi(z,t_f)$ which satisfies Eq. (2.4). In general the application of an electric potential will change the free field energy of the wave function. The change in the free field energy from $t=0$ to $t_f$ is shown in the Appendix to be given by,

$$\Delta\xi_0(0 \to t_f) = \int_0^{t_f} dt \int V(z,t)\frac{\partial J_0(z,t)}{\partial z}dz \tag{2.17}$$

where,

$$J_0(z,t) = q\psi_0^\dagger(z,t)\sigma_3\psi_0(z,t) \tag{2.18}$$

where $\psi_0(z,t)$ is given by (2.6). The quantity $J_0(z,t)$ may be thought of as the current density of the wave function $\psi_0(z,t)$. Recall that $\psi_0(z,t)$ evolves in time according to the free field Dirac equation. Therefore $J_0(z,t)$ will be called the free field current density.

## 3. Hole Theory

The proposition that the negative energy states are all occupied turns a one electron theory into an N-electron theory where $N \to \infty$. For an N-electron theory the wave function is written as a Slater determinant [1],

$$\Psi^N(z_1, z_2, ..., z_N, t) = \frac{1}{\sqrt{N!}} \sum_P (-1)^s P\left(\psi_1(z_1,t)\psi_2(z_2,t)\cdots\psi_N(z_N,t)\right) \quad (3.1)$$

where the $\psi_n(z,t)$ ($n = 1, 2, \ldots, N$) are a normalized and orthogonal set of wave functions that obey the Dirac equation, P is a permutation operator acting on the space coordinates, and s is the number of interchanges in P. Note if $\psi_a(z,t)$ and $\psi_b(z,t)$ are two wave functions that obey the Dirac equation then it can be shown that,

$$\frac{\partial}{\partial t} \int \psi_a^\dagger(z,t) \psi_b(z,t) dz = 0 \quad (3.2)$$

Therefore if the $\psi_n(z,t)$ in (3.1) are orthogonal at some initial time then they are orthogonal for all time.

The expectation value of a single particle operator $O_{op}(z)$ is defined as,

$$O_e = \int \psi^\dagger(z,t) O_{op}(z) \psi(z,t) dz \quad (3.3)$$

where $\psi(z,t)$ is a normalized single particle wave function. The N-electron operator is given by,

$$O_{op}^N(z_1, z_2, ..., z_N) = \sum_{n=1}^N O_{op}(z_n) \quad (3.4)$$

which is just the sum of one particle operators. The expectation value of a normalized N-electron wave function is,

$$O_e^N = \int \Psi^{N\dagger}(z_1, z_2, ..., x_N, t) O_{op}^N(z_1, z_2, ..., z_N) \Psi^N(z_1, z_2, ..., z_N, t) dz_1 dz_2 ... dz_N \quad (3.5)$$

This can be shown to be equal to,

$$O_e^N = \sum_{n=1}^N \int \psi_n^\dagger(z,t) O_{op}(z) \psi_n(z,t) dz \quad (3.6)$$

That is, the N electron expectation value is just the sum of the single particle expectation values associated with each of the individual wave functions $\psi_n$. For example, the free field energy $\xi_0(\Psi^N)$ of the N-electron state is,

$$\xi_0(\Psi^N) = \sum_{n=1}^{N} \int \psi_n^\dagger(z,t) H_0 \psi_n(z,t) dz = \sum_{n=1}^{N} \xi_0(\psi_n(z,t)) \qquad (3.7)$$

**4. Time varying electric potential.**

Assume at time $t=0$, the electric potential is zero and the system is in some initial state which is defined in the following discussion. In HT the unperturbed vacuum state is the state where each negative energy wave function $\varphi_{-1,p}^{(0)}$ is occupied by a single electron and each positive energy wave function $\varphi_{+1,p}^{(0)}$ is unoccupied. The energy of the vacuum state is given by summing over the energies of all the negative energy states. The Slater determinant corresponding to this initial vacuum state can be written as,

$$\Psi^N(z_1, z_2, \ldots, z_N) \underset{N \to \infty}{=} \frac{1}{\sqrt{N!}} \sum_P (-1)^s P\left(\varphi_{-1,p_1}^{(0)}(z_1) \varphi_{-1,p_2}^{(0)}(z_2) \cdots \varphi_{-1,p_N}^{(0)}(z_N)\right) \qquad (4.1)$$

where we assume the following ordering; $|p_1| \leq |p_2| \leq |p_3| \ldots \leq |p_N|$. The total free field energy of the unperturbed vacuum state is then,

$$E_{vac}^{(0)} = \sum_p \varepsilon_{-1,p}^{(0)} = -\sum_p |p| \qquad (4.2)$$

We can add an additional electron provided it consists of a combination of positive energy states $\varphi_{+1,p}^{(0)}$ so that it is orthogonal to the vacuum wave functions $\varphi_{-1,p}^{(0)}$. Let the wave function that defines this positive energy electron, at time $t=0$, be given by,

$$\psi_+(z) = \sum_p f_p \varphi_{+1,p}^{(0)}(z) \qquad (4.3)$$

where the $f_p$ are constant expansion coefficients. Assume that the $f_p$ are selected so that $\psi_+(z)$ is normalized. We can write the Slater determinant of this initial state as,

$$\Psi^N(z_0, z_1, z_2, ..., z_N) \underset{N \to \infty}{=} \frac{1}{\sqrt{(N+1)!}} \sum_P (-1)^s P\left(\psi_+(z_0) \varphi^{(0)}_{-1,p_1}(z_1) \varphi^{(0)}_{-1,p_2}(z_2) \cdots \varphi^{(0)}_{-1,p_N}(z_N)\right)$$

(4.4)

Therefore we have, at the initial time $t=0$, a system which consists of the unperturbed vacuum electrons $\varphi^{(0)}_{-1,p}(z)$ and a single positive energy electron $\psi_+(z)$. Therefore the total free field energy of the system is,

$$E_T(0) = \xi_0(\psi_+(z)) + E^{(0)}_{vac} \quad (4.5)$$

Now we are not really interested in the total energy but in the energy with respect to the unperturbed vacuum state. Therefore we subtract the vacuum energy $E^{(0)}_{vac}$ from the above expression to obtain,

$$E_{T,R}(0) = E_T(0) - E^{(0)}_{vac} = \xi_0(\psi_+(z)) \quad (4.6)$$

which is just the energy of the positive energy electron.

Next, consider the change in the energy due to an interaction with an external electric potential. At the initial time $t=0$ the electric potential is zero and the system is in the initial state given by (4.4). Next apply an electric potential and then remove it at some later time $t_f$ so that,

$$V = 0 \text{ for } t \leq 0; \; V \neq 0 \text{ for } 0 < t < t_f; \; V = 0 \text{ for } t \geq t_f \quad (4.7)$$

Now what is the change in the energy of the system due to this interaction with the electric potential? Under the action of the electric potential each negative energy wave function $\varphi^{(0)}_{-1,p}(z,0)$ evolves into the final state $\varphi_{-1,p}(z,t_f)$. Also the wave function $\psi_+(z)$ evolves into $\psi_+(z,t_f)$. Note that per (4.7) the electric potential is zero at the initial time $t=0$ and the final time $t_f$. Therefore the change in the energy is equal to the change in the free field energy. For the negative energy electrons the change in the free field energy of each electron is,

$$\Delta \varepsilon_{-1,p}\left(0 \to t_f\right) = \xi_0\left(\varphi_{-1,p}\left(z,t_f\right)\right) - \varepsilon_{-1,p}^{(0)} \quad (4.8)$$

and the change in the energy of the positive energy electron is,

$$\Delta \xi_+\left(0 \to t_f\right) = \xi_0\left(\psi_+\left(z,t_f\right)\right) - \xi_0\left(\psi_+\left(z,0\right)\right) \quad (4.9)$$

The total change in the energy of the system is then,

$$\Delta E_T = \Delta \xi_+ + \sum_p \Delta \varepsilon_{-1,p} = \Delta \xi_+ + \Delta E_{vac} \quad (4.10)$$

where,

$$\Delta E_{vac} = \sum_p \Delta \varepsilon_{-1,p}\left(0 \to t_f\right) \quad (4.11)$$

The quantity $\Delta E_{vac}$ is the change in energy of the vacuum. Using these results the energy of the system at $t_f$ with respect to the unperturbed vacuum state is,

$$E_{T,R}\left(t_f\right) = E_{T,R}\left(0\right) + \Delta E_T \quad (4.12)$$

Now we want to evaluate the above quantity. To do this we will use Eq. (2.17). For the vacuum electrons the free field current density $J_0$ is given by,

$$J_0\left(z,t;-1,p\right) = q\varphi_{-1,p}^{0\dagger}\left(z,t\right)\sigma_3\varphi_{-1,p}^{(0)}\left(z,t\right) \quad (4.13)$$

Referring to (2.12) it is evident that $\partial J_0\left(z,t;-1,p\right)/\partial z = 0$. Use this in (2.17) to obtain $\Delta \varepsilon_{-1,p}\left(0 \to t_f\right) = 0$.. That is, the change in the energy of each of the vacuum electrons is zero. Note that result is independent of the applied potential $V(z,t)$. This yields

$$\Delta E_{vac} = \sum_r \Delta \varepsilon_{-1,p}\left(0 \to t_f\right) = 0 \quad (4.14)$$

Next we have to determine the change in the energy of the positive energy electron. The free field current density associated with positive energy electron is,

$$J_+\left(z,t\right) = q\psi_+^{(0)\dagger}\left(z,t\right)\sigma_3\psi_+^{(0)}\left(z,t\right) \quad (4.15)$$

where,

$$\psi_+^{(0)}(z,t) = e^{-iH_0 t} \sum_p f_p \varphi_{+1,p}^{(0)}(z) = \sum_p f_p \varphi_{+1,p}^{(0)}(z,t) \qquad (4.16)$$

Use this in (2.17) to obtain,

$$\Delta \xi_+ (0 \to t_f) = \int_0^{t_f} dt \int V(z,t) \frac{\partial J_+(z,t)}{\partial z} dz \qquad (4.17)$$

Now it is easy to find a state $\psi_+^{(0)}(z,t)$ so that $\partial J_+(z,t)/\partial z$ is non-zero. This can be done by proper selection the expansion coefficients $f_p$. For example let,

$$\psi_+^{(0)}(z,t) = \frac{1}{\sqrt{2L}} \left( \begin{pmatrix} 1 \\ 0 \end{pmatrix} e^{ipz} e^{-ipt} + \begin{pmatrix} 1 \\ 0 \end{pmatrix} e^{ip'z} e^{-ip't} \right) \qquad (4.18)$$

where both $p$ and $p'$ are positive numbers. In this case,

$$J_+(z,t) = \frac{q}{L}\left(1 + \cos\left((p'-p)(z-t)\right)\right) \qquad (4.19)$$

It is evident that the derivate of this quantity with respect to $z$ is non-zero. When $\partial J_+(z,t)/\partial z$ is non-zero it is possible to find a $V(z,t)$ so that $\Delta \xi_+(0 \to t_f)$ is an arbitrarily large negative number. For example let,

$$V(z,t) = -g \frac{\partial J_+(z,t)}{\partial z} \qquad (4.20)$$

where $g$ is a positive number. Use this in (4.17) to obtain,

$$\Delta \xi_+ (0 \to t_f) = -g \int_0^{t_f} dt \int \left( \frac{\partial J_+(z,t)}{\partial z} \right)^2 dz \qquad (4.21)$$

Now the integrated quantity is positive. Therefore as $g \to \infty$ it is evident that $\Delta \xi_+ \to -\infty$. Use this in (4.10) along with (4.14) to obtain,

$$\Delta E_T = -g \int_0^{t_f} dt \int \left( \frac{\partial J_+(z,t)}{\partial z} \right)^2 dz \qquad (4.22)$$

Recall that the energy of the system, with respect to the unperturbed vacuum, at the final time $t_f$ is given by $E_{T,R}(t_f) = E_{T,R}(0) + \Delta E_T$. Now due to the fact that $\Delta E_T$ can be an arbitrarily large negative number then $E_{T,R}(t_f)$ can be negative. Therefore the final energy of the system can be less than that of the vacuum state.

**5. Discussion.**

This result is somewhat surprising. It shows that in HT the unperturbed vacuum state is not the lowest energy state and that it is possible to extract an unlimited amount of energy from an initial quantum state. To review the results of the previous sections we started with an initial system consisting of vacuum electrons in their unperturbed state $\varphi_{-1,p}^{(0)}$ and a positive energy electron $\psi_+$ as defined by (4.3). We then apply an electric potential. The result is that each wave function evolves from its initial state in accordance with the Dirac equation. We find that the change in energy of the vacuum electrons from the initial to final state is zero. This is true for any electric potential. However when we consider the change in the energy of the wave function $\psi_+$ the situation is different. In this case if we set up this wave function so that $\partial J_+(z,t)/\partial z$ is non-zero then we can easily find an electric potential such that the change in energy of the wave function $\psi_+$ can be a negative number with an arbitrarily large magnitude. The net result is that the total energy of the final system is negative with respect to the energy of the vacuum state. This result is consistent with that of previous work [1][2][3].

In the above example the energy of the vacuum electrons doesn't change and the energy of the wave function $\psi_+$, which was originally positive, becomes negative. Now wasn't the Pauli principle suppose to prevent this? What exactly is the Pauli exclusion principle? In the context of HT the Pauli principle is simply the statement that no more than one electron can occupy a given state at given time. Equations (3.1) and (3.2) are the mathematical realization of this statement. The Pauli Principle is a result of the fact that if the initial wave functions in the Slater determinant (see Eq. (3.1)) are orthogonal then these wave functions will be orthogonal for all time. This is a consequence of the fact that the individual wave functions obey the Dirac equation (see Eq. (3.2)). Therefore two electrons cannot end up in the same state.

Therefore the calculations performed in the paper are consistent with the Pauli principle. All the wave functions are orthogonal for all time. This means that the conjecture that the Pauli principle prevents the existence of quantum states with less energy than that of the unperturbed vacuum state is not correct.

**Appendix.**

In this section we will calculate the change in the free field energy of a normalized wave function. Assume at the initial time $t=0$ the wave function is given by $\psi_0(z)$. At some future time $t>0$ the wave function is given by Eq. (2.4). The free field energy of the state at a given time is given by,

$$\xi_0(t) = \int \psi^\dagger(z,t) H_0 \psi(z,t) dz = \int \psi_0^\dagger(z,t) W^\dagger H_0 W \psi_0(z,t) dz \qquad (A.1)$$

Use (2.4) and (2.3) in the above to obtain,

$$\xi_0(t) = \int \psi_0^\dagger(z,t) \begin{pmatrix} -\partial c_1/\partial z & 0 \\ 0 & \partial c_2/\partial z \end{pmatrix} \psi_0(z,t) dz + \int \psi_0^\dagger(z,t) H_0 \psi_0(z,t) dz \qquad (A.2)$$

From this we obtain,

$$\xi_0(t) = -\frac{1}{2q} \int \left( J_0 \frac{\partial(c_1+c_2)}{\partial z} + \rho_0 \frac{\partial(c_1-c_2)}{\partial z} \right) dz + \xi_0(0) \qquad (A.3)$$

where $J_0(z,t)$ and $\rho_0(z,t)$ are the current and charge density, respectively, of the state $\psi_0(z,t)$ and are given by,

$$J_0(z,t) = q\psi_0^\dagger(z,t) \sigma_3 \psi_0(z,t); \quad \rho_0(z,t) = q\hat{\psi}_0^\dagger(z,t) \hat{\psi}_0(z,t) \qquad (A.4)$$

Using the above definitions along with (2.5) and (2.3) we can readily show that,

$$\frac{\partial J_0}{\partial t} + \frac{\partial \rho_0}{\partial z} = 0; \qquad \frac{\partial \rho_0}{\partial t} + \frac{\partial J_0}{\partial z} = 0 \qquad (A.5)$$

Take the derivative with respect to time of (A.3) and use (A.5) to obtain,

$$\frac{\partial \xi_0(t)}{\partial t} = -\frac{1}{2q} \int \left( -\frac{\partial \rho_0}{\partial z} \frac{\partial(c_1+c_2)}{\partial z} + J_0 \frac{\partial^2(c_1+c_2)}{\partial t \partial z} - \frac{\partial J_0}{\partial z} \frac{\partial(c_1-c_2)}{\partial z} + \rho_0 \frac{\partial^2(c_1-c_2)}{\partial t \partial z} \right) dz$$

$$(A.6)$$

Assume reasonable boundary conditions and integrate by parts to obtain,

$$\frac{\partial \xi_0(t)}{\partial t} = -\frac{1}{2q}\int\left(\rho_0 \frac{\partial}{\partial z}\left(\frac{\partial(c_1-c_2)}{\partial t}+\frac{\partial(c_1+c_2)}{\partial z}\right)+J_0\frac{\partial}{\partial z}\left(\frac{\partial(c_1+c_2)}{\partial t}+\frac{\partial(c_1-c_2)}{\partial z}\right)\right)dz \quad (A.7)$$

Use (2.10) to obtain,

$$\frac{\partial \xi_0(t)}{\partial t} = -\int J_0(z,t)\frac{\partial V(z,t)}{\partial z}dz = \int V(z,t)\frac{\partial J_0(z,t)}{\partial z}dz \quad (A.8)$$

Integrate this from $t=0$ to $t_f$ to obtain Eq. (2.17).